# How to Host a Data Competition: Statistical Advice for Design and Analysis of a Data Competition

C.M. Anderson-Cook*, Los Alamos National Laboratory (candcook@lanl.gov)

K. Myers, Los Alamos National Laboratory (kary@lanl.gov)

L. Lu, University of South Florida (lulu1@usf.edu)

M.L. Fugate, Los Alamos National Laboratory (fugate@lanl.gov)

K.R. Quinlan, Pennsylvania State University (krq103@psu.edu)

N. Pawley, Los Alamos National Laboratory (npawley@lanl.gov)

\* corresponding author

Running title: Data Competitions: Strategic Design and Analysis

**Abstract**

Data competitions rely on real-time leaderboards to rank competitor entries and stimulate algorithm improvement. While such competitions have become quite popular and prevalent, particularly in supervised learning formats, their implementations by the host are highly variable. Without careful planning, a supervised learning competition is vulnerable to overfitting, where the winning solutions are so closely tuned to the particular set of provided data that they cannot generalize to the underlying problem of interest to the host. This paper outlines some important considerations for strategically designing relevant and informative data sets to maximize the learning outcome from hosting a competition based on our experience. It also describes a post-competition analysis that enables robust and efficient assessment of the strengths and weaknesses of solutions from different competitors, as well as greater understanding of the regions of the input space that are well-solved. The post-competition analysis, which complements the leaderboard, uses exploratory data analysis and generalized linear models (GLMs). The GLMs not only expand the range of results we can explore, they also provide more detailed analysis of individual sub-questions including similarities and differences between algorithms across different types of scenarios, universally easy or hard regions of the input space, and different learning objectives. When coupled with a strategically planned data generation approach, the methods provide richer and more informative summaries to enhance the interpretation of results beyond just the rankings on the leaderboard. The methods are illustrated with a recently completed competition to evaluate algorithms capable of detecting, identifying, and locating radioactive materials in an urban environment.

1. Introduction

Data competitions, sometimes called machine learning competitions, have attracted considerable attention among the world's community of data and analytics scientists and discipline-specific subject matter experts. This broad involvement provides a model of crowdsourcing for business and government to solve tough high-impact problems in a cost-effective way. Competition hosts often use a commercial platform, such as Kaggle (www.kaggle.com), to hold the competition, rank competitors, and provide a prize (from thousands to millions of dollars) to reward winners. By bringing in new approaches to solving problems, there is potential to accelerate cutting-edge research through the use of data science approaches and the involvement of a more technically diverse set of experts.

Hosting a data competition is time-consuming and often expensive. In this paper we present strategies for designing a competition that will better answer a host's questions of interest, and we build on those design strategies to extract more information from a completed competition than a simple ranking of competitors. These strategies enable hosts to consider more in-depth questions, such as

- Are there fundamental differences between the top competitors?
- What regions of the input space are being solved well, and which poorly?
- Are there changes in competitor rankings for different regions of the input space?

With these planned and implemented approaches, a host can evaluate the alternative solutions and make informed choices for subsequent studies or competitions, thus enhancing their return on investment. Throughout the paper we illustrate our strategies and methods with examples from a data competition we designed and analyzed to evaluate algorithms developed to detect, identify, and locate radiological sources in an urban environment.

There has been limited work in the literature providing guidance or exploring strategies for effectively hosting competitions. Blum and Hardt [8] discuss the overfitting issue with allowing multiple submissions for general competitions. Anderson-Cook et al. [6] describe strategies for building a scalar metric that appropriately balances different aspects of a complex competition problem.

The remainder of the paper is structured as follows: Section 2 describes the typical setup for many data competitions and introduces the urban radiological search example that we refer to throughout the paper. Section 3 describes strategies for creating an effective, impactful competition through the strategic design of data sets and the choice of the scoring metric for ranking competitors. Section 4 describes analysis opportunities enabled by a thoughtful design. These allow a host to gain insights beyond a simple ranking of competitors, and they include both exploratory data analysis and model-based analyses. Finally, Section 5 contains conclusions and discussion.

2. Structure of a Data Competition

The kinds of data competitions we consider here are conducted in a *supervised learning* framework [12]. That is, competitors are provided two sets of data: a *training set*, for which the answers are provided, and a *test set*, for which the competitors will provide their predicted answers for scoring. Competitors develop or "train" their algorithms using the training set, then refine them based on the feedback given by their reported scores on the test set. Typically, the platform that runs the competition will provide a real-time public leaderboard that reflects each competitor's best score and ranking based on their predicted answers for the test set. Competitors can make multiple submissions over the course of the competition, and each is scored and incorporated into the leaderboard if it reflects an improvement over the previous best submission for that competitor. Most competitions impose caps on the maximum daily or total number of submissions.

Within the test set, there is a further division of the data that is not disclosed to the participants. A fraction of test data forms the *public test set*, which is used to score and rank the competitors on the public leaderboard while the competition is running. The remaining test data form the *private test set*. The final score for each competitor is based on the private test set and is not shared with the competitors until the competition closes. The private leaderboard, based on the scores on the private test set, specifies the final ranking and winners of the competition.

The host has the flexibility to specify what data comprises the training, public test, and private test sets, often subject to a practical limit on the total amount of available data. The host also chooses a static evaluation formula or *scoring metric* to define the score of each submission and its ranking on the public and private leaderboards. It is our understanding that the vast majority of current data competitions rely exclusively on the leaderboard to evaluate and rank the submitted solutions. To provide timely, succinct feedback on competitors' performance during the competition, the scoring metric is usually a simple scalar summary that quantifies the accuracy and effectiveness of a solution for solving the competition task(s). This metric, when properly defined, encompasses the key aspects of the problem under investigation with the competition, and seeks to identify its top solutions. However, there are several potential limitations to this approach: First, by necessity, the scoring metric is created before the competition opens. Hence anything that the host learns by observing competitor contributions cannot be incorporated into revisions of the metric. Second, the leaderboard summary is a global number that amalgamates responses across a large number of instances, each of which could represent different regions of the problem space. One solution might be best in one region of this space, while another might be superior elsewhere. When the host wants to choose a solution for a particular sub-problem, understanding different relative performance could lead to different choices that should be tailored to the individual question to be answered. Finally, since many data competitions involve multiple tasks, the scoring metric for the leaderboard must combine evaluation of all of these tasks and may be too simplistic to allow deeper understanding of the relative performance of the different solutions and address multiple questions of interest to the host. How to weigh the different contributions through penalties for incorrect portions of the answer can have a profound effect on how different competitors are ranked.

The post-competition analysis described in Section 4 below allows uncoupling of the different aspects of the problem as well as detailed comparisons between competitor solutions throughout the problem space. Since the constraint of summarizing and ranking the competitors with a single summary has been relaxed, exploratory data analyses and model-based characterization of the solutions can be used to describe patterns in the solutions. The post-competition analysis can be made more powerful and informative by having a well-chosen set of data for the competition as described in Section 3. Anderson-Cook et al. [6] describe strategies for selecting an intentional coherent test set for data arising from simulated or measured sources. By combining a strategic construction of the test set with a detailed analysis, it is possible to maximize the information gained and value from hosting a competition.

> **Case Study: Urban Radiological Search.** Throughout this paper, we provide concrete examples of the strategies and methods in the context of a competition to detect, identify, and locate radiological sources in an urban environment, [datacompetitions.lbl.gov/competition/1/](datacompetitions.lbl.gov/competition/1/). This competition used simulated measurements mimicking those collected by a radiation detector being driven along typical urban streets. The simulations were performed at Oak Ridge National Laboratory, where they could flexibly simulate data for a wide variety of scenarios. The inputs for these scenarios were chosen to mimic the diversity of urban environments seen in practice.

A key feature of urban radiological search is being able to separate the background signal (generated from benign emitters of radiation, like buildings and pavement, in the urban environment) from a localized source. We divided the input factors into several categories: characteristics of the background, characteristics of the sources, and characteristics of the detector's movement. For the background factors, several versions of urban streets were used with different configurations and compositions for the buildings and features.

For the source factors, we considered five different radioactive source types, plus an additional source defined as a combination of two of the sources. These sources include weapons grade materials and isotopes common in medical or industrial settings:

1. HEU: Highly enriched uranium
2. WGPu: Weapons grade plutonium
3. $^{131}$I: Iodine, a medical isotope
4. $^{60}$Co: Cobalt, an industrial isotope
5. $^{99m}$Tc: Technetium, a medical isotope
6. A combination of HEU and $^{99m}$Tc

The other source factors included its location on the street, its strength, and whether it was shielded in a dampening container. With close engagement from the subject matter experts, we combined the location, strength, and shielding factors into a measure of the signal-to-noise ratio (SNR), which is used as a descriptor in the analysis later.

For the detector factors we considered its speed in meters per second as it traveled along the street, the traffic lane of travel, and the starting / ending points within a street.

The data were generated using a stochastic simulation code developed at Oak Ridge National Laboratory. Each "run" or instance of data in the training and test sets was specified by selecting values for more than 100 parameters. The individual file sizes for each run ranged from 160 KB (when the detector is moving quickly and over a shorter section of road) to 7.3 MB (moving slowly over a longer path with more active background). To keep downloads and manipulation of the data manageable for the competitors, we constrained the total file size for the zipped data (training and test sets together) to 10 GB. This served as a "budget" for the number of runs to be included in the combined training and test data sets.

The urban radiological search competition was open to competitors working at or affiliated with U.S. government laboratories and ran from February through May 2018. Sixteen teams participated, and each team could be comprised of multiple individual participants. Across all of the teams there were nearly 1000 total submissions, with the top four teams contributing between 100 and 250 submissions each. In the examples that follow we focus primarily on the top three teams: LBNL (Lawrence Berkeley National Laboratory), LANL-W (one of two teams from Los Alamos National Laboratory) and Python Hacks (one of four teams from Lawrence Livermore National Laboratory).

3. Design Considerations for Data Competitions

Like designed experiments offer more cost-efficient strategies to simultaneously change multiple input factors for studying the underlying relationship between the input and response variables, strategically designed competition data can offer more efficient information for accelerate improvements and drive better solutions from hosting data competitions. With a limited size of the competition data, simply using a collection of raw data that satisfy the size constraint could result in a loss of opportunity. For example, if the goal of a competition was to find an algorithm that offers the highest success rate for classifying some rare event of interest, using the typical data that are severely unbalanced may have preserved the raw data features but could fail to make good use of the valuable competition resources and lose the opportunities to develop more efficient machine learning algorithms for detecting cases of interest. For a competition with the goal of finding a general solution suitable for a broad scenarios of interest, using convenient data from localized regions or narrow time windows or specific scenarios that are hard to generalize to broader scenarios might easily lead to competitors solving a constrained or even wrong problem. We think the goal of the competition should not just be to find the best solution for the particular data set provided to the competitors, but rather to identify best approaches to the general class of problems that the host intends to solve. This should drive the process for hosting the competition, and help with making choices about which data to choose (what to include and exclude) to encourage the competitors to build robust and general solutions.

While perhaps obvious to state, the competition should have a clearly specified and articulated goal against which data can be compared to evaluate if a competition has potential to achieve success. A critical part of success of the competition is to have clear objectives for what a desirable participant solution should be able to do, and ensuring that the available data are adequate to match the goals, and that the provided data sets are informative and effective for driving the best sustainable solutions.

Here we describe six strategies for designing competition data sets to drive the maximum outcome of the competition:

1. Encourage competitors from diverse technical backgrounds.
2. Select data that adequately covers the region of interest.
3. Emphasize data of maximum interest.
4. Discourage algorithms from overfitting to idiosyncrasies in the data.
5. Adapt standard design of experiments principles to the competition scenario while preventing competitors from exploiting unintended artifacts in the data.
6. Create a leaderboard to balance the goals of the competition, and appropriately reward the most desirable performance characteristics.

We discuss each strategy in more detail below and illustrate their implementation in our urban radiological search competition.

### 3.1. Encourage competitors from diverse technical backgrounds

One of the advantages of data competition or crowdsourced solutions is that they can draw from a larger candidate pool of experts than might not normally be involved. In order to take advantage of the opportunity to include a technically diverse set of competitors, we think it is beneficial to think about what information should be provided to allow those who are new to the subject area to gain traction. For instance, specialized information that subject matter experts traditionally use to solve the problem should be included. This will allow each discipline's competitors to leverage and build upon their current state-of-the-art tools. In addition, potential obstacles to participation, such as downloading the data, understanding its format, and making submissions, should all be carefully thought out to minimize the overhead for competitors to get started. If making the first submission to enter the competition has too steep a learning curve, potential participants might be lost.

> For the urban radiological search competition, we provided basic information about radiation detection data. This included providing plots and data illustrating spectra for each of the sources measured in a vacuum, both with and without shielding. This allowed competitors with data science backgrounds and no expertise in or experience with radiation detection data to quickly make progress. The competition website included information about file formats, and we limited the total size of the zipped data to 10 GB to make it easier for competitors with limited computational resources to participate.

*3.2. Select data that adequately cover the region of interest*

As with traditional design of experiment (DoE) [11, 22], matching the design region to the problem of interest is essential for designing an effective, informative competition that answers the right questions. This entails identifying the factors to be varied, the appropriate ranges of each of the factors, and potential constraints on viable factor combinations that may make the region irregular. As with traditional DoE, subject matter experts (SMEs) typically begin with a larger number of candidate factors, and then downselect to identify those thought to be most influential. We suggest, where possible, to begin with an available "superset" of candidate data considerably larger than the training and test sets that the competitors will ultimately receive.

This strategy may change for a data competition because of constraints on available data. Where the competition builds on simulated data, the capability of the data generator may restrict which regions are available. If the currently available data sets are too limited to span the space of interest, we think it is helpful to invest resources to expand the available superset of data to improve the ability of the competition to answer the real aims of the host.

> For the urban radiological search competition, the initial version of the Oak Ridge simulation model allowed generation of detector data for a vehicle moving in a single direction down a fixed-length multi-block stretch of a single street. When we examined an initial superset of data, similarities between runs were quite strong. The initial data sets essentially asked the competitors to develop an algorithm to answer the question "Can you detect a source on this particular street?", when the desired goal was to develop methods to detect a source on any American urban street. By focusing on the overall goal of seeking a general solution, the Oak Ridge team redesigned their simulation to allow considerable additional flexibility and variability in the generated data.

> While the SME investment made to enrich the simulation was substantial, their efforts enabled the competition to answer the true question of interest. As a consequence, the quality of solutions developed by the competitors may have been substantially higher.

*3.3. Emphasize data of maximum interest*

After defining the factors of interest and their appropriate ranges, we need to choose instances from throughout that space to form the training and test sets. However, not all regions of the space have equal value for answering the questions of interest. We want instances that are sufficiently challenging to push algorithm development. We also want a collection of instances that can effectively highlight the differences between solutions. Some instances might be trivially simple while others are impossibly hard. Data sets that lead to all of the competitors getting the same answer (all right, or all wrong) are an inefficient use of resources. The sweet spot for providing the most informative data to the competitors is in the middle range, with sufficient challenges but also a good possibility of getting the right answer if the algorithms are sufficiently capable.

For simplicity, consider a binary classification problem where the competitors are asked to predict a 1 or a 0 for each instance in the test set. A logistic model based on the levels of the input factors can be used to model how these different input factors drive changes on the correct classification rate. With a range of anticipated algorithm performance, traditional design selection and optimization strategies can be leveraged for selecting more informative data for the training and test sets. If we further simplify and just consider a single input with a known relationship to the probability of correct classification, the D-optimal design [17] that offers the most precise estimation of model parameters under the assumed relationship places half of the points at the location with probability of success 0.176 and the other half at the location with probability of success 0.824 [1].

However, data competitions often have a number of important complicating factors that preclude the use of this simple design strategy:
- We have multiple inputs, so this is inherently a high dimensional space for which we want to understand the relationship between inputs and our questions of interest.
- Often, we have multiple questions that we wish the competitors to answer.
- We have multiple competitors, each with potentially multiple submissions, who will have solutions that perform differently across different regions of the input space for each of the different questions.
- Perhaps most importantly, we do not know a priori what the algorithms will look like and how well they will be able to solve the different aspects of our competition.

Hence we think the goal for creating the data sets should be to provide sufficient data in the regions of interest that we anticipate will allow for good estimation of performance for each of the competitors near the top of the leaderboard.

Quinlan et al. [20] and Quinlan & Anderson-Cook [19] propose strategies for creating designs based on two or more prior distributions for the anticipated probability of correctly answering a single question of interest. We suggest using two priors to bound the space of performance of interest. One of the priors quantifies current capability for the best available algorithm before the competition. This is assumed as the lower bound for the performance of interest, since the competition hopes to inspire improvement beyond that capability. The second prior is the dream capability that might be achievable by the top competitors by the end of the competition, which can be specified with the help from the SMEs.

Based on these two priors, the goal is to select a range for each factor that allows us to distinguish between competitors with high probability. In high dimensions, we select the most interesting range for each factor separately. Values of the factor for which the priors suggest P(success) > 0.824 are unlikely to yield many failures, and values for which P(success) < 0.176 are unlikely to currently yield many successes. If values in the ranges between these lower and upper bounds are emphasized more in the competition data, then we increase the likelihood of good estimation of performance and meaningful comparisons between competitors.

Of course, eliciting appropriate priors in a high dimensional input space from SMEs can be challenging. Based on our experience, a good strategy for choosing the lower bound is to run the best available algorithm on the entire superset and use this to calibrate the difficulty level of the problem. Identifying a useful upper bound is more challenging, but using the current best model of performance can provide a baseline for expectations of improvement from the new algorithms. While this may provide an imprecise estimate of performance, it at least can help to rank the relative difficulty of different regions. In addition, SMEs may be able to provide some insights about physical limitations of what any algorithm might be able to solve, and these could serve as a proxy for that dream performance prior at the upper end of difficulty.

> For the urban radiological search competition, we had multiple questions for the competitors: detection, identification, and location of different sources. Since identification was considered the most important of the three questions, we focused primarily on this for defining the most interesting region. Our superset included one partition for each source type plus another for runs containing no source. We used the best available detection and identification algorithm on our entire superset of data. Within each partition, we used the results from the algorithm to fit logistic models as a function of background level, detector speed, and (for the source partitions) source strength and shielding. Using this model, we determined regions in the input space that were sufficiently difficult for this current algorithm to justify their inclusion in the training and test sets.
>
> In addition, we consulted with SMEs to frame the region where they thought that an exceptional algorithm might be able to discern a signal, both for detection and identification. This was used as an upper bound. To reduce dependence on this as a prior, and since the range of interest for each of the inputs had been determined separately, we continued to use the entire range for each of the inputs, weighting the more promising regions more heavily. At the conclusion of this phase, the entire region of the input space was still represented, with some regions emphasized more heavily than others to reflect their anticipated relative importance to understanding and comparing competitor solutions.

*3.4. Discourage algorithms overfitting to idiosyncrasies in the data*

One of the obstacles to using data competitions to develop long-term solutions for complex problems is the required static nature of the data sets. In order to have fair comparisons between competitors and for them to understand the requirements of the solution, the training and test sets remain unchanged throughout the competition. At the same time, the competition structure allows the competitors to repeatedly submit answers for the static test set to improve their algorithm performance. From the competitor perspective, this provides opportunities for learning from their previous submissions and experimenting with adjustments to the solution algorithms. If the training and test sets share similar performance characteristics – say in the case where the host randomly assigns available data to the training or test set -- then competitors can improve the leaderboard score by increasing the complexity of their models to capture the idiosyncrasies found in the training data.

From the host perspective, these fixed data sets could potentially lead to competitors solving the wrong problem, especially if the training and test sets share artifacts that are unique to the competition data and not to general scenarios of interest. Then even the winning solution is likely to be ineffective when used on a more general problem. Ultimately, the host wants the winning solution to perform well, not only in the competition setting, but also in new (perhaps currently unanticipated) scenarios.

The potential risks of model overfitting based on a single data set are well documented [12], but this problem is exacerbated because of the repeated submission aspect of data competitions [8]. Hence, we think it is important to use the construction of the training and test sets as a way of mitigating overfitting and encouraging extrapolation to unexplored regions. The useful practice of subdividing the test set into a public and private component as described in Section 2 presents an opportunity for implementing this mitigation.

To force competitors to handle new scenarios well, we think it is helpful to construct the training, public test, and private test sets with increasing levels of difficulty. If the training set excludes the most difficult scenarios that the competitors will be scored against, their algorithms will need to be able to adapt for solving new challenges first presented in the public test set. In addition, since the private test set will ultimately determine the winner, we think it is helpful to include new scenarios that the competitors could not tune their algorithms to through multiple submissions against the public test set. In this way, the private test set data provides a good proxy for assessing how algorithms might be expected to perform on more general scenarios of interest.

Figure 1 shows a two input variable "cartoon" to demonstrate our strategy for differentiating the various data sets to help prevent overfitting. The bottom left corner represents the easiest corner of the input space, while values of the inputs moving to the right or top increase the level of difficulty. We begin with the private test set in yellow, which ultimately determines the competition winner. We include the entire region of interest in the private test set since we want to estimate performance throughout. We choose the public test set, where competitors receive feedback on their submissions, to be a sub-region of the entire space. The most difficult portions of the range for some or all of the factors are removed or dramatically undersampled. In addition, "holes" in the space are created where no data are included for a portion of some factor ranges. Finally, the training data, where both data and answers are provided, is a further subset of the public test set space. Holes are again incorporated, some of which correspond to those between the public and private test sets (shown as "yellow holes" in Figure 1), and some distinct to the training and public test sets (shown as a "blue hole").

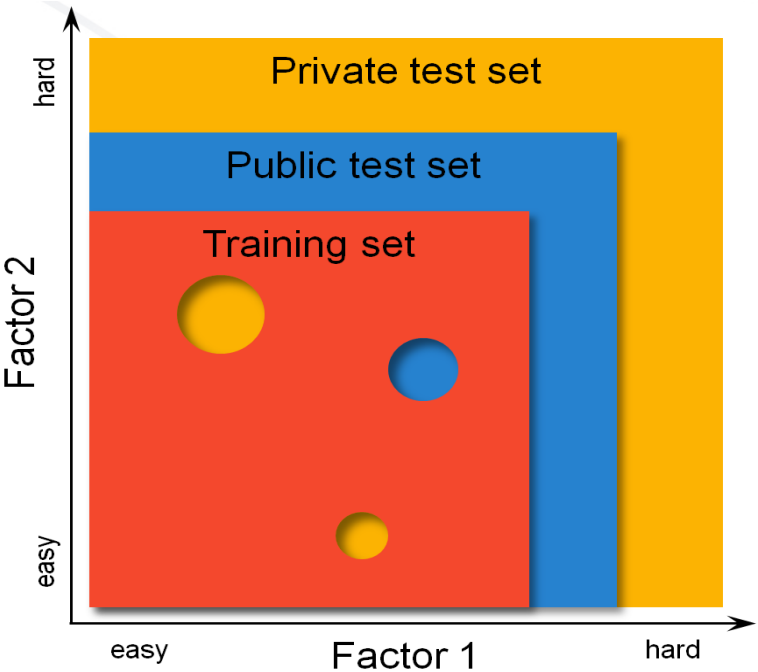

Figure 1: Relationship between training, public test, and private test data for a 2D input space.

The reasoning behind this strategy is to force the competitors' algorithms to demonstrate the ability to handle both interpolation and extrapolation. Based on the training data, competitors have not seen answers in the more difficult region in the public test set (blue perimeter outside of red training set), and hence their solutions need to demonstrate the ability to expand into new scenarios. This process is repeated for the stretch from the public to private test sets, where the host can see how well the algorithms can again extend to further new scenarios. The interpolation elements (holes) provide an opportunity to check if there are differences in local behavior that can be a symptom of overfitting.

To ensure a fair competition, we recommend clearly describing that there are differences between the training, public and private data sets to the competitors. This can be done in the instructions or description of the competition dataset available for almost all data competitions. Making this information available can help the competitors calibrate their expectations on the competition tasks, reinforce the idea of solving the general problem of interest and avoid overfitting to the particular training data that are provided, and also help them choose appropriate strategies for seeking the best solution that the host desires.

Taking sub-ranges of multiple inputs can quite rapidly diminish the amount of data available to assign to the training and public test sets. How much to reduce the ranges should be balanced with data availability when populating these data sets.

> For the urban radiological search competition, beginning with a superset 10 times the size of the final competition data sets allowed for aggressive divisions into training, public test, and private test sets. We first considered the extrapolation part of specifying the data for the factors speed, shielding, and background. Detecting a source becomes more difficult at higher speeds. Similarly, shielding a source makes it more difficult to detect or identify. Some of the background conditions were also known to make the search more difficult. For each of these factors, with an anticipated gradient from easy to hard, we specified sub-ranges from which to draw the training and public test sets.

> For the interpolation aspect, we considered the multiple configurations of the street and the multiple locations for placing a source. While the street configurations produced different patterns of background, they were considered effectively of the same difficulty. Hence, only half of the street configurations were shared in the training set, ¾ of them included in the public test set, and all of them included in the private test set. Similarly, ½, ¾, and all source locations were used in the three data sets, respectively.

*3.5. Adapt standard design of experiments principles to the competition scenario*

In this section we consider principles from traditional design of experiments [16,17] that we translated into strategies for effective hosting of a data competition. A key distinction from typical experimentation is that we consider the competitors to be adversaries who are trying to leverage any information communicated through the data sets. This includes information both intentionally and unintentionally shared by the host. In most other experiments, the choice of factor combinations is dictated solely by the goals of the experiments or constraints. Hence some design principles, such as replication, design balance, and randomization, need to be considered through fresh eyes. We discuss each principle below.

**Replication.** Since many competition responses are categorical (e.g., detect/no detect or which source type in our example), having replicates can dramatically improve the power of a logistic model to quantify performance throughout the region. If the simulator has a stochastic component, then creating replicates at the same nominal input conditions should result in differences between the replicates. Nuisance factors (those factors which make the data look different to the competitor, but are not thought to meaningfully impact the difficulty or nature of the problem) can be used to make replicates look less similar. However, current file utilities make comparisons between files straightforward, so including instances too similar to each other may be of limited value since this is might be detected by savvy competitors.

> For the urban radiological search competition, the duration of the detector moving along the street was treated as a nuisance factor, since this duration was not of interest to the subject matter experts, and was not thought to impact the difficulty of the problem. Hence manipulating it to vary the look of the replicates did not compromise our ability to estimate performance. In this way replicates of a scenario could be used, where the stochastic nature of the simulation generating code and the selected starting and ending points of the street made the runs look quite different to the competitors.

**Design balance.** Typically in designed experiments, a goal is to have perfect or near-perfect balance between levels of each factor. That is, we typically include the same number of instances of each level of a factor. This allows for better orthogonality between factors and independent estimation of their effects, as well as similar standard errors, which make comparisons between the magnitudes of effect more straightforward.

For the design of a competition, balance needs to be tempered by unequal emphasis on different regions of the design space as described earlier, as well as an interest in avoiding unintended artifacts that competitors can leverage when tuning their solutions. For competition questions that involve identifying different categories of responses, the goals of the competition should dictate choices that are made. For example, consider the trade-off between false positives (sounding an alarm when there is nothing to respond to) and false negatives (missing an actual event). These generally have very different associated costs, and so the competition test sets should be designed to provide adequate precision for estimating each of these rates.

For the urban radiological search competition, we treated the design for each source type (6 source types plus no source) as separate mini-experiments, which were constructed independently and then combined at the end. Balance, in particular for the number of runs for each source, was a feature that the competitors could potentially exploit. The balance between the number of runs containing a source and those containing no source was also something that needed to be managed between the data sets. In an actual urban radiological search setting, the number of "no source" runs would (we hope) vastly outnumber those "with source", but there is no requirement to have the competition data set mimic reality based on this aspect. We wanted to have adequate information to be able to precisely assess the false positive rate in different regions of the input space. So we decided on the fraction of no source vs. with source runs for the public test set and for the private test set, and then divided the "with source" runs among the 6 different sources. We intentionally did not balance the number of runs for each source. We estimated the minimum number required to estimate the relationship between inputs and the response for a given source, and then used this as the number of runs for the easiest source. The remaining sources were assigned slightly more runs, to avoid too much balance between sources. In addition, we intentionally had different fractions of runs for each source between the public and private test sets.

**Randomization.** In a traditional experiment, randomization serves the purpose of protecting against unknown systematic effects during the running of the experiment. This seems less critical in a simulated data set, since it is much less likely that there will be lurking factors impacting the results. However, there are still good reasons to randomize the order of files for the public and private test sets. Since the training data set contains the complete set of answers, there is no reason to randomize the order of instances within it. In fact it may be helpful for the competitors to group all runs within a category together for ease of understanding patterns.

For our test data, we inter-mixed the public and private subsets in the test data to complicate probes by the competitors to identify which runs are associated with different aspects of the test set. We also randomized the order of source and no source runs.

*3.6. Create a leaderboard to reflect the goals of the competition*

The scoring metric provides a single formula chosen before the launch of the competition for ranking the competitor solutions from best to worst on the leaderboard. As the competitors making their best efforts to improve their leaderboard scores during the competition, this scoring metric will drive the direction of the competition and competencies of the resulted solutions. If this ranking is not strategically chosen to match the goals of hosting the competition, then competitors could focus on aspects of the problem that are of lesser importance, and/or the overall winner might not be the one that provides the most desired solution to the competition. By intentionally matching the goals of the competition to the construction of the leaderboard, the host has control to appropriately emphasize the different aspects of the competition.

In the urban radiological search competition, "no source" runs (scored with 1 for correct classification or 0 otherwise) were weighted half as much as "with source" runs, since they represented answering a simpler question. For the "with source" runs, the multiple different criteria were combined into a single score with an additive desirability function [9]. The three components of detect, identify, and locate were combined with an additive form:

$$\text{Score}_{\text{source}} = w_{\text{det}}\text{Det} + w_{\text{iden}}\text{Iden} + w_{\text{loc}}\text{Loc} \quad \text{with} \quad \sum w_{\bullet} = 1$$

We consulted with the SMEs to determine initial weights for each of the components that were thought to reflect competition priorities. Then we constructed various mock competitor submissions with systematic errors – e.g., a submission that answered every test set run correctly except for those containing highly enriched uranium (HEU) sources, or a submission that answered every test set run correctly but got the location wrong by 1 second – and computed the leaderboard for all the mock submissions using those initial weights. We then made changes to the weights to ensure that the rankings best matched the SMEs' priorities for a winning submission. Anderson-Cook [3] describes this approach to spot-checking alternative weightings to obtain a desirable robust ranking of top contenders.

4. Post-Competition Analyses

The design portion of hosting a data competition is critical for its success, since providing the right data enables quality answers to the questions of interest. Having a robust analysis strategy to consider the different questions of interest allows for greater knowledge and understanding to be extracted from the investment of the competition. Indeed a good design enables a rich analysis.

Here we discuss both exploratory data analysis and model-based analyses to support better understanding of the input space, comparisons between solutions, and evaluation of the chosen leaderboard metric. We discuss each approach in turn.

*4.1. Exploratory data analysis*

Exploratory data analysis of each submission can provide useful summaries that provide preliminary information about performance and team-to-team differences. Global summaries about the fraction of correct answers for different portions of the data can give an indication of overall performance.

*Contingency tables* [2, 7] are effective descriptive summaries for categorical responses. A *confusion matrix* [11, 12] is a contingency table with two dimensions, typically with the true answers listed as rows and the predicted answers as columns. Along the diagonal are the counts or proportions of instances that a competitor predicted correctly, and the off diagonal entries show instances where the competitor was "confused" by the problem. Note that the amount of data for different questions of interest is likely to be unbalanced, for instance due to intentional choices made during the data selection process to avoid unintended artifacts in the data. In addition to global summaries, it is helpful to explore differences in the results for different input values. Graphical summaries that connect the responses for individual runs can be connected to the inputs used to create the data.

The summaries illustrated in this section consider just the raw data and presented different subsets of it to gain preliminary understanding of fundamental differences between the results obtained by different teams. This can be helpful for probing areas of deeper exploration with more formal approaches.

In the urban radiological search competition, there were several aspects of interest. For runs where a source is present: (a) detecting the presence of the source, (b) identifying which of the 6 possible sources it is, and (c) locating where along the path the source is placed. For runs where no source is present: (d) correctly stating that no source is present. We initially focus on the detection and identification portions of the competition and categorize the answers for each of the runs. Since each team could enter multiple submissions, we begin the exploration by using the results from the best scoring submission for each team.

Figure 2 shows a confusion matrix summary for both detection and identification for the winning team's (LBNL) best submission as scored against the test set data (both the public and private test sets combined). Rows correspond to the true state while columns indicate what the competitor specified as their answer. Figure 2(a) shows the conditional detection probabilities for the entire test set data, with the first row showing that the probability of detecting a source when a source (S) is present is approximately 74.7% (peach color), while the probability of correctly saying no source (NoS) is present is 93.2% (orange color). Figure 2(b) shows the conditional identification probabilities, with information about how each of the sources was classified among the 7 possible choices. Clearly the diagonal shows that the most common choice for each source is the correct identification. The lighter blue color in the last column shows that each of the sources is sometimes missed and called a no source run. Source 6 (which is a mixture of sources 1 and 5) is most frequently misidentified as source 1.

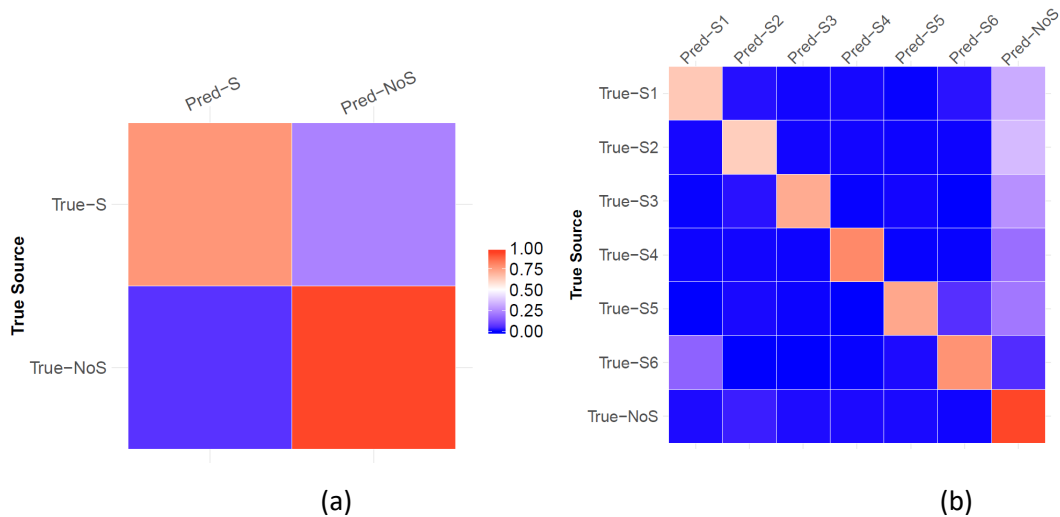

(a) (b)

Figure 2: Confusion matrix summaries for LBNL's best submission for (a) detection and (b) identification. S and NoS mean "source present" and "no source present" respectively.

The summary in Figure 2 provides a global look across all the runs. Because the runs are constructed by specifying different input choices, it is also possible to examine how the correct and incorrect answers are distributed in key input regions. For example, Figure 3 shows the spread of the results for the winning submission on the portion of the private test data from weapons grade plutonium (WGPu) with shielding. (We consistently use WGPu with shielding for the purpose of illustration.) Similar plots could be constructed for each of the sources, and for any subset of the input space, for complete exploration of the results. Other key inputs were chosen for the x-axis (the speed of the detector in meters per second) and y-axis (the signal to noise ratio or SNR, which summarizes the strength of the source signal relative to the background noise). In the plot, green indicates correct detection and identification, orange is correct detection and incorrect identification, and red means that the source was not detected. Based on this plot, it is clear that the most difficult regions are for low SNR, and there is some increase in difficulty as the speed of the detector increases. The clustering of the orange points for low speeds and small SNRs suggests that this is a region where detecting a source is possible, but correctly identifying it is challenging. Also, the relatively small number of orange points suggests that if WGPu is detected, it is most often correctly identified.

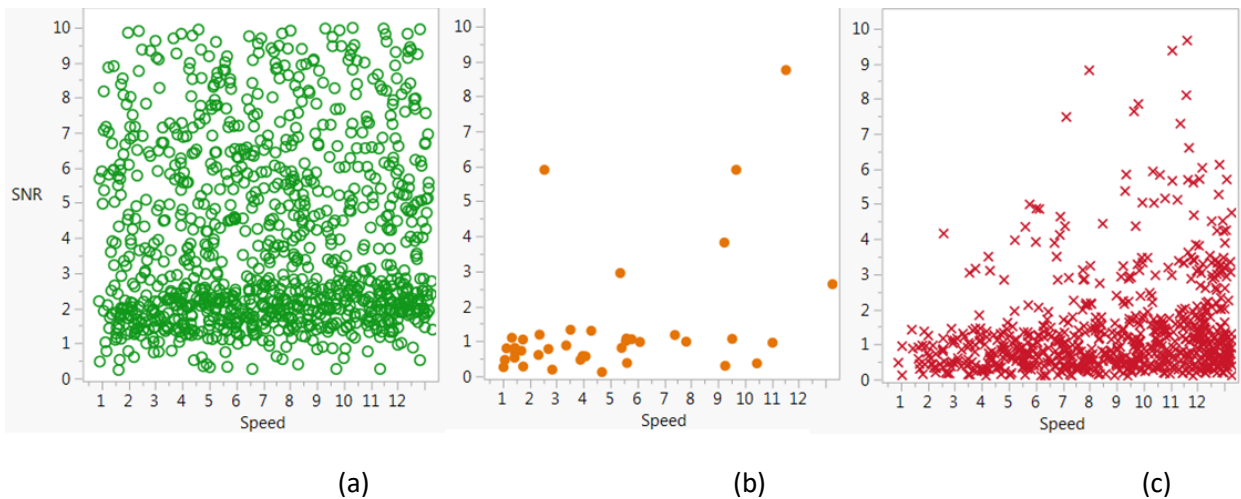

(a)             (b)             (c)

Figure 3: Scatterplot of identification for the best submission from team LBNL for data based on runs with weapons grade plutonium (WGPu) present and with shielding. Here we consider the parameters detector speed (in m/s on the x-axis) and source signal-to-noise ratio (SNR, y-axis). (a) green = correct detection and identification; (b) orange = correct detection, incorrect identification; (c) red = source not detected.

We can also make comparisons between teams. Each panel in Figure 4 provides information about the subset of data for one scenario (each source or no source), broken out by correct identification (dark green), correct detection with incorrect identification (dark blue), or no source detected (light blue). Each bar in each panel represents the performance of one of the teams, sorted from highest overall score to lowest. From this plot, we are able to extract some general trends from the data. First, highly enriched uranium (HEU) and WGPu are generally harder to detect or identify than iodine ($^{131}$I) and cobalt ($^{60}$Co) as seen by the relatively larger light blue regions. Secondly, for most sources, if teams were able to detect the source, then they could also correctly identify it (as seen by the small dark blue regions for most source types). The exception to this is Source 6, which was a mixture of HEU and technetium ($^{99}$Tc). Here the detection rate was high (better than for any other source), but not surprisingly there were greater problems with correctly identifying that source.

If teams had all used fundamentally similar approaches, we might expect that the results would monotonically decrease from left to right in a given panel in Figure 4, as the best teams were able to solve the problem a bit better than teams with lower scores. However, we do see that some teams were able to do better at detection or identification than the best team for some sources. These differences between teams and different sub-categories of the data provide incentive to explore more deeply to understand differences between teams' solutions. If teams could solve different parts of the overall competition with varying degrees of success, then this may represent an opportunity to leverage the best of each solution into a "super-solution" that can dramatically outperform any individual team's results.

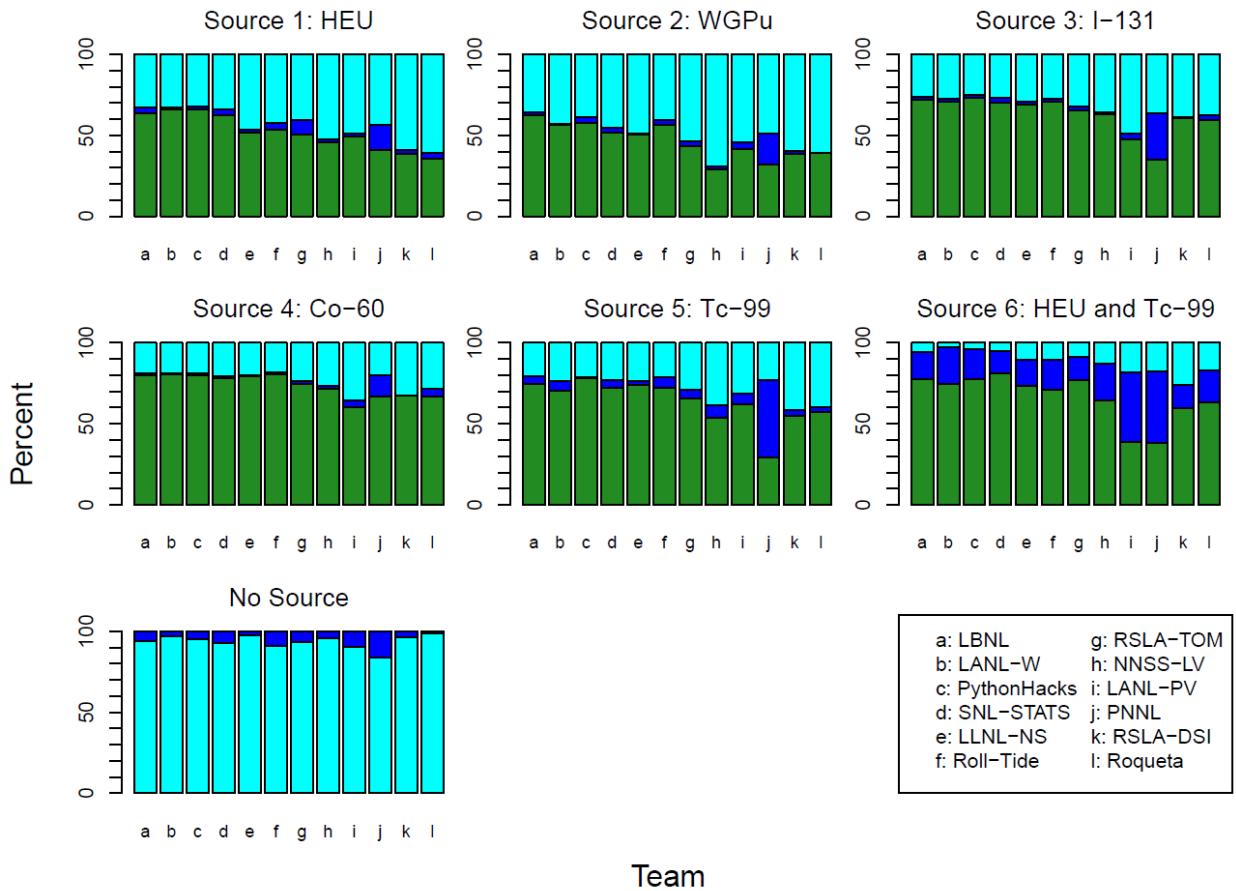

Figure 4: Barplots of correct identification, detection, and no source found for each source for the best submission for each of the top 12 teams. Green indicates correct detection and identification, dark blue shows correct detection and incorrect identification, and light blue indicates that no source was found.

Table 1 and Figure 5 show some initial exploration of the different strengths and weaknesses of the solutions of the top two teams. In Table 1 the rows correspond to correct detection and identification (I), correct detection with incorrect identification (D), and incorrectly missing the source entirely (X) for the top team, LBNL. The columns show the same information for the second best team, LANL-W. The diagonal entries show where the teams reached consistent results, while the off-diagonals identify differences in algorithm performance. In general, we anticipate that the larger the fraction of off-diagonal entries, the greater the potential differences in the approaches used for their solutions.

Table 1: Summary of differences between top submissions from 1st (LBNL) and 2nd (LANL-W) place teams for WGPu. I = correct detect and identification, D = correct detection, but incorrect identification, and X = incorrectly missed source. Bold entries indicate where the two teams obtained the same results. The first entry is the cell count, while the entry in parentheses indicates the fraction of the total number of WGPu runs.

|  |  | Team 2: LANL-W | | | |
|---|---|---|---|---|---|
|  |  | I | D | X | Total |
| Team 1: LBNL | I | **930 (0.504)** | 4 (0.002) | 134 (0.073) | 1068 (0.578) |
|  | D | 4 (0.002) | **1 (0.001)** | 34 (0.018) | 39 (0.021) |
|  | X | 33 (0.018) | 10 (0.005) | **697 (0.377)** | 740 (0.401) |
|  | Total | 967 (0.524) | 15 (0.008) | 865 (0.468) | 1847 (1) |

When we compare the top two teams, we see that LBNL correctly identified slightly more cases (1068 vs. 967) and correctly detected slightly more (39 vs. 15). However, it is interesting to note that LANL-W was able to correctly identify 37 cases that were incorrectly identified by LBNL (rows D and X of column I), with 4 of those being detected by LBNL, and 33 having been missed completely. Seeing that nearly 12% (219 of 1847) of the WGPu cases lie on the off-diagonals (4+134+34+4+33+10 = 219) reveals potential differences in the teams' approaches, even though the overall scores for the teams were quite close.

To dig more deeply into where these differences lie, a variation of the scatterplot shown in Figure 3 can be considered. Figure 5 shows the WGPu data for just the scenarios where shielding was included, with Figure 5(a) showing data corresponding to the diagonals in Table 1. Figures 5(b) and (c) show the off-diagonal cases where LBNL outperformed LANL-W and vice versa. From this figure, some patterns can be explored. For example, LBNL seemed to be better than LANL-W at detecting WGPu for low SNR and slow speeds (cluster of pink points in bottom left corner of Figure 5(b)). If there are strong clusters in regions of the input space, then this may represent opportunities to compare solution approaches to leverage improved performance of one algorithm over the other.

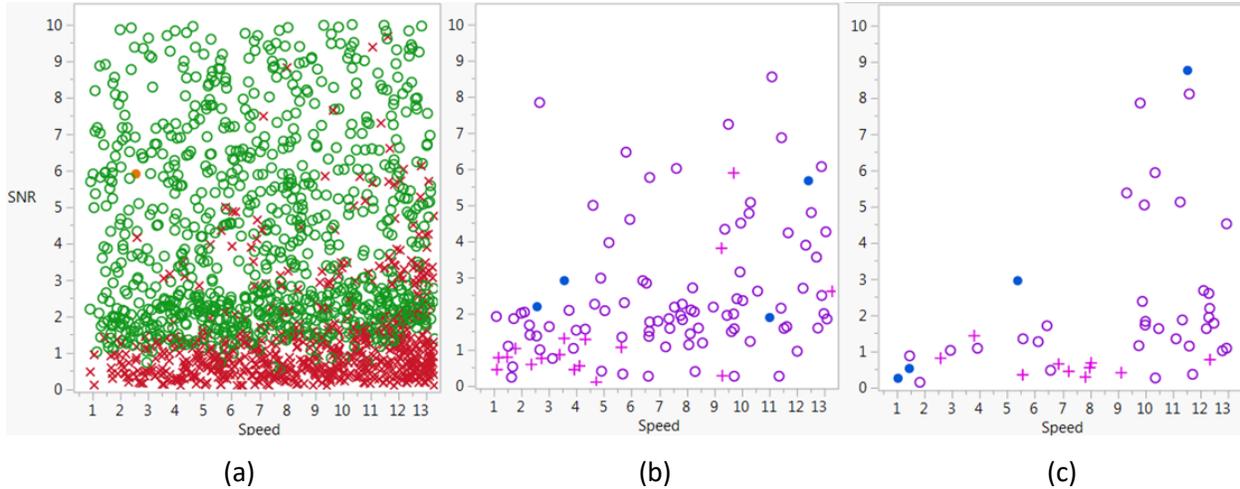

(a)                          (b)                          (c)

Figure 5: Scatterplot of comparisons between performance of top 2 teams (LBNL and LANL-W) for WGPu with shielding. (a) Results where teams match (green open circle = correct detection and identification, orange filled circle = correct detection but incorrect identification, red X = missed source). (b) Runs where LBNL outperformed LANL-W (top-right cases in Table 1) with blue filled circle = LBNL correctly detected and identified, LANL-W correctly detected but misidentified; purple open circle = LBNL correctly detected and identified, LANL-W missed source; pink + = LBNL detected but misidentified, LANL-W missed source. (c) Runs where LANL-W outperformed LBNL, with same color coding as (b) but with teams reversed.

### 4.2. Model-based analyses

As the analysis of the different competitor solutions evolves, the methods used to gain understanding become more formal. In the early stages, looking at the raw data and outcomes can show differences at a high level, while later delving into more detail can reveal more subtle comparisons between the solutions of different competitors. An underlying theme in the post-competition analysis is the goal of exploring whether competitors used different approaches to solve the problem, and whether these differences can be exploited to perhaps develop a super-solution which leverages the best of each of several algorithms to achieve even higher performance.

Using a model to summarize the relationship between the input factors and the results of different sub-questions of the competition can provide a more formal way to gain a deeper understanding of patterns in the data. For categorical variables, such as the detection and identification portions of the urban radiological search competition, a generalized linear model (GLM) [15, 18] can be helpful. For continuous variables, such as the size of the miss for the location portion of the urban radiological search competition, a standard linear model for characterizing a response surface [17] can provide insights.

GLMs are flexible extensions of linear regression models to handle broader distributions of the response variable than just continuous responses well modeled by the Normal distribution. GLMs allow the response variable to follow any distribution in the exponential family, which includes the commonly used Normal, Binomial, and Poisson distributions, and also allow the mean response to depend on the explanatory variables through a link function (e.g., logit, probit, or log). GLMs are popular models for categorical responses. For example, Poisson regression models are often used for modeling count data. The logistic or probit regression models are popular choices for modeling binary response variables. For example, if $Y$ represents a binary response which follows a Bernoulli distribution with $p = \Pr(Y = 1)$, then the logistic regression model links the expected mean response (i.e. the proportion), $p = E(Y)$, to the exploratory variables, $x$, through the logit link function as given by $\text{logit}(p) = \log\left(\frac{p}{1-p}\right) = x'\boldsymbol{\beta}$.

For the urban radiological search competition, the test set data were partitioned into 7 separate subsets, one for each of the six source types and one for the no source data. To understand both the estimated probability of detection and probability of identification for each source, the following logistic models were fit:

$$P(\text{detection}) = \frac{e^{x'\beta}}{1+e^{x'\beta}} \quad \text{and} \quad P(\text{identification}) = \frac{e^{x'\beta}}{1+e^{x'\beta}}$$

where

$$x'\beta = \beta_0 + \beta_{SNR}X_{SNR} + \beta_{Shield}I_{Shield} + \beta_{Speed}X_{Speed} + \beta_{Background}I_{Back} + \beta_{Lane}X_{Lane}$$

for the input factors signal-to-noise ratio (SNR - continuous), shielding (indicator variable 0/1), speed (continuous), background (indicator variable for one of multiple) and traffic lane (closest or furthest from source - continuous for one of 4 lanes). We performed careful model selection to obtain the best fitted model for each team based on its top submission for each source. We first considered a full model including all the relevant input variables and their higher order terms and interactions for all the scenarios to allow for greater flexibility in the modeled shape of the response surfaces. Then we sequentially removed non-significant higher order terms from the full model to avoid overfitting and reduce the variability of the prediction for each individual scenario. We performed a lack-of-fit test between the full and the reduced model to check the adequacy of the simplified model. By removing some of the spurious terms from the model, the overall variance of the model was reduced, leading to improved interpretability and less risk of overfitting to idiosyncrasies in the data. We fit these models to the top scoring submission from each of the top teams to obtain tailored estimations of individual team performances across the different source types.

In terms of gross similarities across the different input factors, the signal-to-noise ratio, speed, and shielding were highly significant, while the background layout and the lane were rarely significant. This was unintuitive to the SMEs who helped to design the simulations used for the competition, as they had anticipated that being further from the source would make it more difficult to detect it. The differences in the layout of the background captured by $I_{Back}$ were not thought to fundamentally change the level of difficulty of the problem to be solved, but were primarily used to obfuscate the patterns of the background. Figure 6 shows contour plots for the highly significant input factors for different cases of the WGPu scenario for the top team, LBNL. We also found that the three-factor interaction was rarely significant. Among the two-factor interactions, the interactions between SNR and shielding and between SNR and speed were significant more often than the interaction between speed and shielding across all explored scenarios and competitors.

Examining the results from Figure 6, we note some patterns in the estimated surfaces. First, the probability of detection (top panels) is equal to or higher than the probability of identification (bottom panels) for all regions of the input space. The difference between the two surfaces at a given location corresponds to the size of the "correctly detect and incorrectly identify" group in Figure 3. Second, when we compare the left and right plots in a given row, we see that the "with shielding" scenario was generally harder than "no shielding". Third, the LBNL team was able to have very high probability of detection and probability of identification for cases when the SNR ratio was large and for slower speeds of the detector (upper-left regions of each panel). There is a large proportion of the input space where the probability of both detection and identification is greater than 90%, indicating that for these cases, it is quite likely that LBNL's solution can adequately solve the problem of interest. This matches the qualitative results that were observed in Figure 3 with few errors in detection or identification in the top portion of the plot.

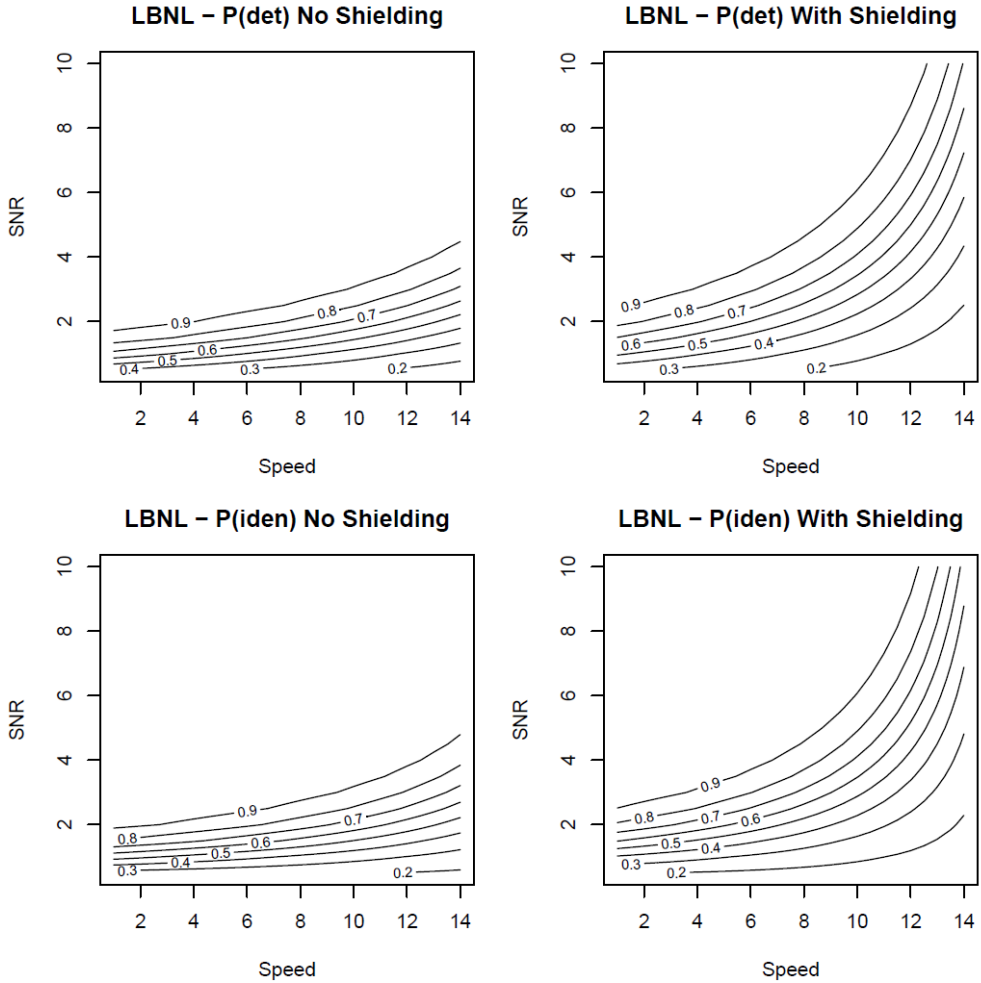

Figure 6: Contour plots of the probability of detection (top panels) and the probability of identification (bottom panels) for WGPu based on the results from the best submission from the top team, LBNL. The left and right panels of each row are estimated from the same model, but shown for $I_{Shield}$ = 0 and 1, respectively.

Next, we can also examine the relative impact of the different factors on the two probabilities. Since the contours are more horizontal for the "no shielding" case, we can infer that changes in the SNR have a greater impact on the capability of detection or identification than changes in the speed of the detector. For the "with shielding" case, the contours are more curved with similar changes in the vertical and horizontal directions. This suggests that for the shielded case, changes in SNR and speed have similar impacts on the probabilities of detection and identification.

If there is an interest in assessing which regions of the input space look to be well-solved and those for which there is still need for improvement in the solution, it is helpful to look more broadly than just the top team. We can use our model-based analyses to compare teams.

Recall from Table 1 and Figure 5 that there were runs which the second place team, LANL-W, was able to answer correctly while the top team was not. Hence, Figure 7 shows the contour plots for the top 2 teams for the WGPu with shielding scenario to compare their performance. The general shape of the two sets of contours look similar, with some small differences in the curvature of each contour. However, based on these results, it does look like for SNR values greater than 5 and speeds less than 6 (top left corner), both algorithms have a very high probability of correctly identifying WGPu.

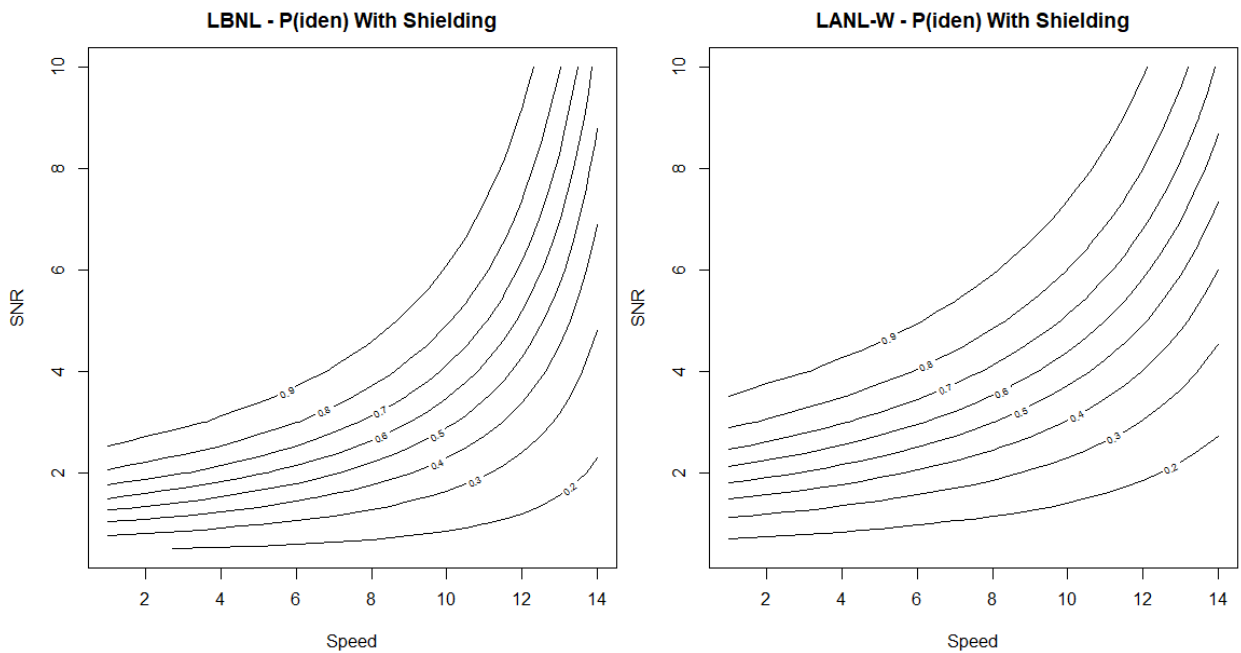

Figure 7: Contour plot of the probability of identification for WGPu with shielding for the top two teams.

In addition, if we examine the bottom right corner, we can gain some understanding about how challenging the different algorithms find the low SNR and high speed cases. Neither teams exceeded a 20% chance of identifying the source for these most difficult cases. This matches with what we saw in Figure 3 for the top team with very few runs with correct identification in this corner of the input space.

For the runs with sources, the additional component of the competition, besides detection and identification, was to see how well the competitors could locate where the source was placed in the urban environment. In this case, a standard linear regression model could be used to examine patterns. The response of interest was formulated as the size of the miss from the true source location in seconds, where misses before and after the true location were treated as equivalent. Hence a model of the form

$$|\text{time miss}| = \beta_0 + \beta_{SNR}X_{SNR} + \beta_{Shield}I_{Shield} + \beta_{Speed}X_{Speed} + \beta_{Background}I_{Back} + \beta_{Lane}X_{Lane} + \varepsilon$$

was used where the factors considered were the same as for the probability of detection or identification. We assume that the error terms, $\varepsilon$, are independent and identically distributed. Again, results for this response followed similar patterns across all of the sources. The SNR, speed and shielding factors were highly significant with increases in the size of the miss increasing for more difficult versions of each factor. This time, in addition to these factors, the traffic lane was also moderately significant for some of the sources, such that being in the lane furthest away from the location of the source increased the size of the miss. Overall, this portion of the competition was generally less difficult for the competitors and there were fewer distinctions between teams. In other words, if the competitors could detect and identify the source, then pinpointing its exact location was not that challenging.

For the runs with no source, the response of interest was the probability of correct classification as a run with no source. Since many of the input factors included previously were functions of the source, the modeling for this response was considerably simpler. Again, a GLM was used where the relevant input factors to explore were the speed of the detector and differences in the background configurations. We fit a model of the form

$$P(\text{correct classification}) = \frac{e^{x'\beta}}{1+e^{x'\beta}} \quad \text{with} \quad x'\beta = \beta_0 + \beta_{Speed}X_{Speed} + \beta_{Background}I_{Back}$$

for each competitor's best submission and compared the results. Note the linear component of the GLM $x'\beta$ can include higher order terms for related explanatory variables, and we performed model selection to choose the best fitting model for each team. For our example, the background variable was almost never significant for modeling the probability of correction classification for the no source case for the top winning teams. Both the linear and the quadratic terms of the speed were significant for the top 2 teams, LBNL and LANL-W. Figure 8 shows the estimated probability of correct classification for the no source scenario for these teams with the associated estimated uncertainty as measured by confidence intervals for the curves. It is interesting that the probability of correct classification generally rises as the detector moves at higher speeds. One explanation is that those runs don't provide enough data (because the simulated detector isn't on the street for a long enough time) for competitors to confuse a background element for an actual extraneous source.

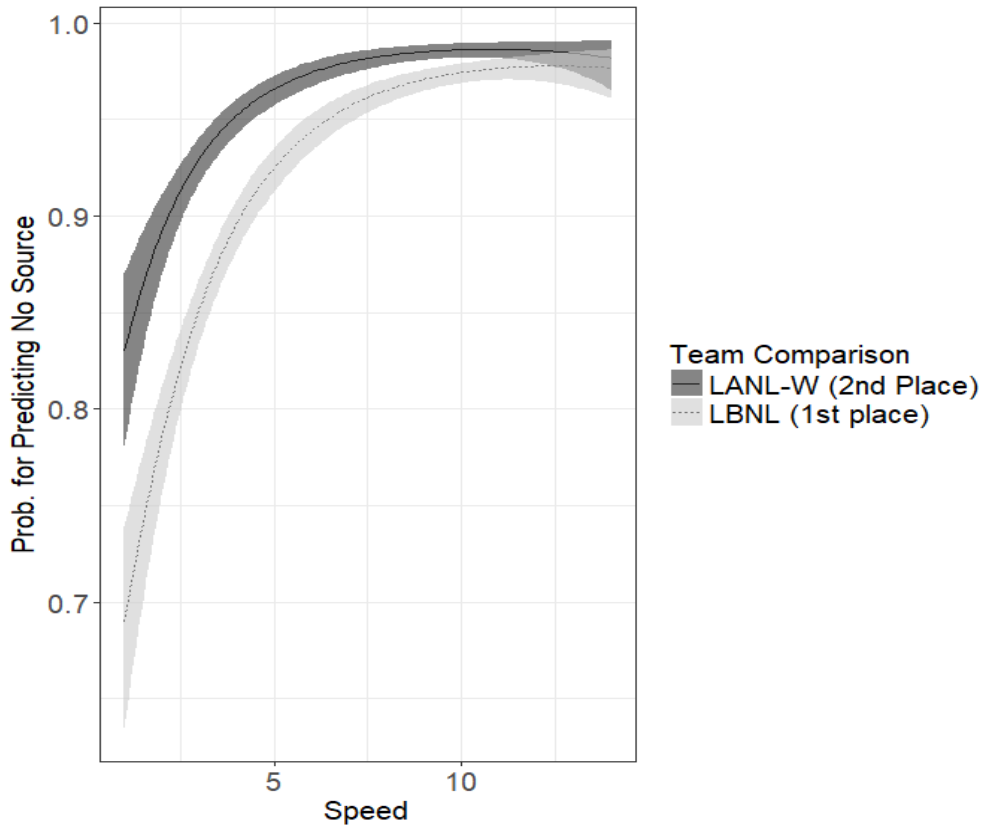

Figure 8: The probability of correct detection (true negative) for no source cases for the top 2 teams with the estimated uncertainty (the 95% pointwise confidence bands).

As we compare the top two teams, we see that the winning team does not have the best performance on correctly classifying the no source scenario. The second place team, LANL-W, has consistently higher estimated probability of correct classification than the winning team. With the large size of the competition data, the estimates have pretty good precision with narrow pointwise confidence bands for both curves. The two teams perform similarly at very high speed values (above 12 meters per second) since it becomes too hard to detect anything when the detector is moving really fast.

Through the use of GLMs and standard regression, it is possible to develop response surface models that describe the general patterns of each team's solution. From this exploration is it possible to gain understanding about which of the input factors play a role in influencing the quality of the solution, to quantify relative impact, and to visualize where regions of excellent, fair, and poor performance exist.

We can also use model-based analyses to quantify differences between competitor results for different input regions. This could reveal opportunities for combining complementary approaches to create an even stronger hybrid algorithm. To make precise where these opportunities lie, we now illustrate using our models to find regions in the input space with differences in performance.

In Table 1 and Figure 5 we summarized an exploration of where the top two teams performed similarly and differently. Here we consider this question more formally using the GLMs estimated above. Figure 7 shows contour plots for the probability of correct identification for the top two teams for the scenario of WGPu with shielding. While the gross patterns of the surfaces seem similar, it can be challenging to identify where there are differences in the probability of identification.

Figure 9 highlights the differences between the fitted models for the top two teams (LBNL vs. LANL-W) for the probability of identification for WGPu with shielding. Values of zero in this plot (light blue regions) correspond to the two surfaces having the same probabilities, while positive (darker blue) regions show where LBNL has a higher probability of identification than LANL-W. We use a p-value approach from hypothesis testing of the two surfaces having the same mean to evaluate the statistical significance of the observed difference considering the estimation uncertainty. In Figure 9, different symbols are used to indicate different significance levels: white solid circles for p-values less than 0.01, black solid circles for p-values between 0.01 and 0.05, and black open circles for p-values between 0.05 and 0.1. Since multiple comparisons are made across a number of input locations across the input space of interest, we use a Bonferroni approach [16] to adjust the p-value for conducting multiple simultaneous comparisons by multiplying the p-value of a point-wise comparison by the number of input locations evaluated (i.e., the number of comparisons that are considered). This approach can be quite conservative when a large number of input locations are evaluated across the input space. If improved power is desired for the simultaneous test, then a less conservative approach should be used for making the multiple comparison adjustment, or a coarser grid of input locations could be used for the evaluation. Figure 9 clearly shows that the largest positive difference between the two surfaces (LBNL outperforming LANL-W) is found for SNR values close to 2 and slower speeds. What is also identified, that was more difficult to see from Figure 7, is that there is a region for high speeds and larger SNR values where the LANL-W team is slightly outperforming the LBNL team (yellow region). However, this is not statistically significant based on the Bonferroni simultaneous test.

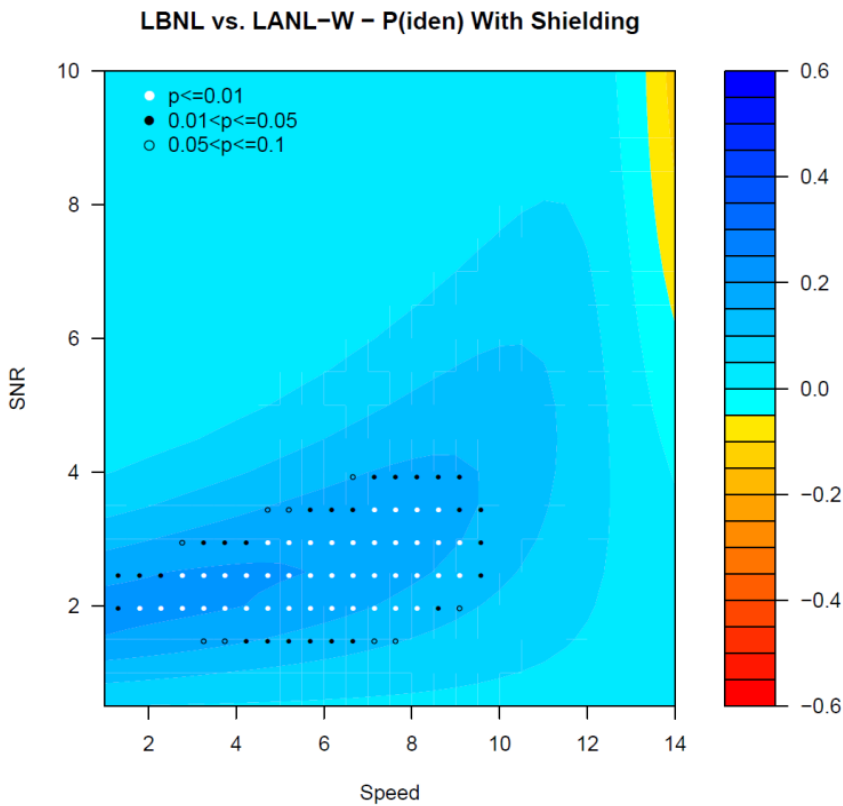

Figure 9: Contour plot of the difference in the mean estimated surfaces for the probability of identification for WGPu with shielding between the top two teams: LBNL vs. LANL-W. Positive values (in blue shades) indicate that LBNL has a higher probability of identification, while negative values (in yellow-orange-red shades) indicate better performance by LANL-W. The white and black solid circles and black open circles indicate input locations with significant differences with p-values being less than 0.01, between 0.01 and 0.05, and between 0.05 and 0.1, respectively. We adjust for multiple comparisons with the Bonferroni approach.

One concern of the p-value approach is that with a large sample size it is easy to declare a statistical significance for a small amount of difference which might not be of practical importance to the problem of interest. Equivalence testing [4, 25, 26] has been developed for evaluating the practical equivalence between groups. Equivalence testing is still a p-value based approach that relies on a formal test. Stevens and Anderson-Cook [23, 24] describe a more flexible strategy for characterizing the differences between two response surfaces based on fitted logistic regression models. Although originally developed for a reliability context, the described probability of agreement is suitable for examining if the differences between the fitted surfaces are of practical importance. The question of whether different populations should be treated as practically equivalent is one that occurs in a wide variety of applications. For the purposes of this exploration, formal testing is not necessarily needed, but some quantification of how likely two surfaces are to overlap could be beneficial.

A key difference between traditional hypothesis testing and the probability of agreement is how the initial assumption about the two surfaces is stated. Traditional hypothesis testing assumes that the surfaces to be compared are equivalent until there is evidence to the contrary. Probability of agreement, on the other hand, starts with the assumption that the surfaces are different until there is sufficient evidence to assert that they are practically equivalent [21]. Another key difference is that the user has the option to specify what a meaningful difference between the surfaces is for each application. The probability of agreement can be used as an alternative or complementary summary to the difference plot (Figure 9) for evaluating the practical importance of the observed differences between teams.

> By comparing the estimated response surfaces for the different responses in the urban radiological search competition, we are able to more precisely identify opportunities where further exploration into the competitors' approaches might lead to advantages and the potential to improve the overall solution. It is worth noting that the top two teams were formed by experts from quite different fields. The top team, LBNL, mainly had radiation detection experts, while the second place team, LANL-W, was made up of five statisticians. Learning the strengths and weaknesses of the two completely different sets of skills could increase the opportunities to leverage strengths of different techniques for creating improved solutions.

5. Discussion and Conclusions

In this paper, we have presented some strategies for both the design and analysis of data competitions. Because the effort and financial investment of hosting a competition is substantial, being intentional about what data to present to competitors is critical for getting good return. We think selecting data which provide maximum ability to distinguish between the performance of different submissions while stimulating growth of solutions is key to taking advantage of the opportunity that a competition offers. We also think it is important to be inclusive for a breadth of different technical audiences, and to construct a leaderboard metric which reflects the priorities of the most desirable solution. By focusing on matching the data to where competitor solutions are sought, it is more likely to avoid rewarding the wrong characteristics of the solutions.

While the leaderboard is an essential part of evaluation of the competitors' submissions in a data competition, it must be simplistic and scalar by necessity to meet the real-time unique ranking requirements. We propose that a more detailed post-competition analysis can be highly beneficial to maximize what is learned through the competition. One of the goals of this analysis is to understand if there is potential for the best of each team's submission to be combined into a global solution that surpasses all of the individual contributions. Initially starting with exploratory data analysis methods can provide basic insights about where top competitors were able to perform well, and also observed differences between results from different teams. To make these comparisons more formal, model-based analysis such as generalized linear models or standard regression models can be used to characterize the relationships between the inputs and the responses with response surfaces. For binary categorical responses, such as for correct classification, detection, or identification, logistic regression can provide a simple model form for easy interpretability and comparison. For continuous responses, linear models are useful. These models allow assessment of the relative importance of different factors on the quality of the answers and also facilitate the construction of response surfaces and visualization of the relationships. Tools such as difference plots and formal evaluation of these differences can compare solutions to understand where they are effectively the same, and where they are potentially complementary.

In the urban radiological search competition, the post-competition analysis showed a number of useful results about where in the input space the top teams could solve the questions posed well, and where additional work is still needed. It also revealed some differences between the top teams and showed that the second place team might have better local performance in some regions of the input space for answering a certain sub-question than the winning team. While the numerical and graphical summaries shown throughout the paper were based on our example, their general nature should allow them to be easily adapted to other competitions where the objectives may be different. Our goal in presenting the example was not to provide an exhaustive list of summaries to be used in competitions, but rather to illustrate the potential for decomposing the leaderboard scoring metric into more detailed components for which the right summary can be developed.

We anticipate that many competition hosts will not be involved in just a single competition, but rather a sequence of competitions that evolve from each other. For this purpose, the post-competition analysis has an additional benefit of providing insights about where to focus future data, both for comparison of the algorithms that were developed in each competition, and to explore regions where further development of solutions is still needed to meet the host's needs.

There are areas where additional research would be beneficial, including examining the robustness of the leaderboard scoring metric to the subjective weighting of multiple objectives of a competition, such as the urban radiological search competition's interest in the three components of detection, identification, and location of potential sources. A Pareto fronts approach [5, 17], which identifies non-dominated solutions for simultaneously optimizing multiple objectives and explores trade-offs and robustness of solutions to different user priorities, could provide insights into how the host valued the individual components of the scoring, and how this affected the overall ranking.

When a sequence of competitions is planned, the improved understanding from the post-competition analysis of a previous competition can help calibrate the leaderboard scoring metric for the later competitions. There are also opportunities to provide other carefully designed in-competition summaries to the participants which could foster accelerated improvements to their algorithms without inadvertently alerting them to the structure of the particular data used in the competition. As competitions become increasingly prevalent, it will be beneficial to have a suite of analysis methods that provide the right tools to enable the host to make the most of their financial and effort investment.